\documentclass[runningheads]{llncs}
\usepackage[T1]{fontenc}
\usepackage{amsmath}
\usepackage[table]{xcolor}
\usepackage{booktabs}
\usepackage{multirow}
\usepackage{makecell}
\usepackage{float}

\usepackage{graphicx,verbatim}

\begin{document}

\title{Automatic Patient-Specific Microwave Ablation Planning Accelerated by a Physics-Guided Deep Learning Model}

\titlerunning{Automatic MWA Planning with Physics-Guided Deep Learning}

\author{Seonaeng Cho\inst{1} \and
Minjee Seo\inst{1} \and
Minju Seol\inst{1} \and
Juil Park\inst{2} \and
Joon Ho Kwon\inst{2} \and
Kyungho Yoon\inst{1,3}\thanks{Corresponding author}}

\authorrunning{S. Cho et al.}

\institute{School of Mathematics and Computing (Computational Science and Engineering), Yonsei University, Seoul, Republic of Korea\\
\email{yoonkh@yonsei.ac.kr}\and
College of Medicine, Department of Radiology, Yonsei University, Seoul, Republic of Korea \and
Innovative \& Intelligent Computational Science Institute (IN2CSI), Seoul, Republic of Korea
}
  
\maketitle              
\begin{abstract}

Microwave ablation (MWA) is a promising minimally invasive treatment for liver tumors, but its therapeutic outcome strongly depends on patient-specific planning of antenna insertion trajectory, power, and treatment duration. Accurate numerical simulation can provide physically reliable ablation predictions; however, its high computational cost limits its use in optimization-based planning, where repeated forward evaluations are required. To address this issue, we propose a digital twin-based automatic planning framework that combines a neural ablation prediction model with a genetic algorithm. The model was trained on multiphysics simulation data generated from patient-specific tumor and vessel structures, antenna configurations, and treatment conditions, and was used as a fast forward model during planning. The prediction model achieved a Dice score of 95.1\%, enabling accurate deep learning-based optimization. In 13 unseen planning cases, the proposed method improved ablation efficiency by 54.3\% and reduced organ damage by 55.0\% compared with clinician-defined planning, while slightly shortening the insertion path length by 3.3\%. Most generated plans were also judged clinically applicable by MWA specialists. Furthermore, the framework enabled approximately 420-fold faster planning than numerical-simulation-based planning, demonstrating its potential as a fast digital twin for quantitative and personalized MWA treatment planning. The code is available at: {https://github.com/SeonAengCho/MWA-Planning.git}


\end{abstract}
\section{Introduction}

Microwave ablation (MWA) is a minimally invasive thermal therapy in which an antenna is inserted into the body to deliver microwave energy and induce thermal necrosis of tumor tissue \cite{simon2005microwave}. Compared with surgical resection, which may be limited by general anesthesia, bleeding risk, and prolonged recovery, MWA can be performed through a percutaneous or small-incision approach \cite{yang2025impact,gu2025microwave,marrero2018diagnosis}. Owing to these advantages, MWA has been increasingly adopted for the treatment of solid tumors, including liver tumors.
\newline \indent The therapeutic outcome of MWA, however, strongly depends on preoperative planning. Insufficient ablation may leave residual tumor tissue and increase the risk of local recurrence, whereas excessive ablation may damage surrounding normal tissues or critical organs \cite{tan2021initial,kim2010minimal}. Therefore, preprocedural planning is essential to determine the antenna insertion position and trajectory, applied power, and treatment duration while accounting for the location, size, and shape of the tumor as well as nearby vessels and organs \cite{li2025multi,zhang2019computer}.
\newline \indent In current clinical practice, MWA planning is often performed based on two-dimensional ultrasound or computed tomography (CT) images. Such image-guided planning makes it difficult to fully understand the three-dimensional spatial relationship between the tumor and surrounding anatomical structures, as well as the true tumor volume and morphology \cite{an20203d,liu2013three}. In addition, treatment parameters are frequently selected based on manufacturer-provided protocol tables or empirical rules. These approaches do not sufficiently account for patient-specific tumor geometry, vascular structures, or heat-sink effects \cite{van2023computational}. As a result, treatment planning remains highly operator-dependent and may not provide a systematic strategy for individualized therapy.
\newline \indent To overcome these limitations, patient-specific three-dimensional numerical simulations have been investigated to predict treatment outcomes \cite{radjenovic2021efficacy,frackowiak2023first,heshmat2024using}, enabling simulation-based treatment planning \cite{gao2019conformal,nahmed2026digital}. However, MWA simulation is a computationally expensive multiphysics problem that involves electromagnetic analysis, bioheat transfer, and thermal damage estimation. This computational burden becomes particularly problematic in optimization-based planning, where forward simulations must be repeatedly evaluated for various treatment parameters.   
\newline \indent Recently, studies have sought to reduce this computational burden by training neural networks to approximate complex numerical simulation processes, thereby enabling faster inference \cite{seo2024multi}. In thermal therapy, Convolutional neural network (CNN)- or Transformer-based image models have been used to predict temperature distributions and ablation zones \cite{shin2024physrfanet,cho2025towards}, and neural operators have also been used to account for heat-sink effects \cite{meister2022fast}. In MWA, recent studies have explored the prediction of ablation zones from CT images and treatment conditions \cite{keshavamurthy2024pre}.
\newline \indent In this study, we propose a digital twin-based automatic MWA planning framework that combines a neural ablation prediction model with a genetic algorithm (GA). Ablation regions were generated using numerical multiphysics simulations based on patient-specific tumor and vessel structures, antenna configuration, power, and treatment duration. These data were used to train an affine transformation-based multimodal neural network, which was then used as a fast forward model within the optimization loop. The genetic algorithm searched for treatment plans on the patient-specific 3D body twin while considering feasible antenna trajectories, organ avoidance, normal tissue preservation, target coverage, and safety margin coverage. Through this framework, we propose a digital twin-based automatic MWA planning method with three main contributions: 1) personalized planning that reflects patient-specific tumor, vessel, and major anatomical structures, 2) reduced computational cost via neural network-based forward prediction, and 3) clinical applicability validated through quantitative comparison with clinician-defined planning and assessment by MWA specialists.

\section{Method}

\subsection{Simulation-based Training Data Generation}

\subsubsection{Data acquisition}

We used the HepaticVessel dataset from the Medical Segmentation Decathlon \cite{antonelli2022medical}, which provides abdominal CT images with liver tumor and intrahepatic vessel labels. A total of 60 patient CT datasets were used to train and test the neural ablation prediction model. Tumor and vessel masks were separated, resampled to 1-mm isotropic spacing, and cropped into $81^3$ volumes centered on each tumor centroid. For each tumor, 500 input conditions were randomly generated, including a six-dimensional antenna position and direction vector $(x, y, z, dx, dy, dz)$ and a two-dimensional treatment vector consisting of power and duration. The antenna position was sampled within the tumor, while power and duration were sampled from $45$--$100~\mathrm{W}$ and $60$--$300~\mathrm{s}$, respectively.

\subsubsection{Numerical simulation}

Numerical simulations were performed for 30,000 generated input conditions to obtain the ablation regions. The simulation consisted of three multiphysics steps \cite{berjano2006theoretical,gorman2022numerical}: electromagnetic field calculation using the Helmholtz equation, tissue temperature estimation using the Pennes bioheat equation, and cellular necrosis estimation using the Arrhenius model. The Helmholtz equation was solved using the finite element method (FEM), whereas the Pennes bioheat equation was solved using the finite difference method (FDM). The vessel region was maintained at $37^\circ\mathrm{C}$, with a convection boundary condition applied at the vessel--tissue interface. Finally, the Arrhenius damage integral was accumulated over time, and voxels with a final value greater than 1 were defined as the necrotic region.

\subsection{Neural Ablation Prediction Model}

\begin{figure}
\includegraphics[width=\textwidth]{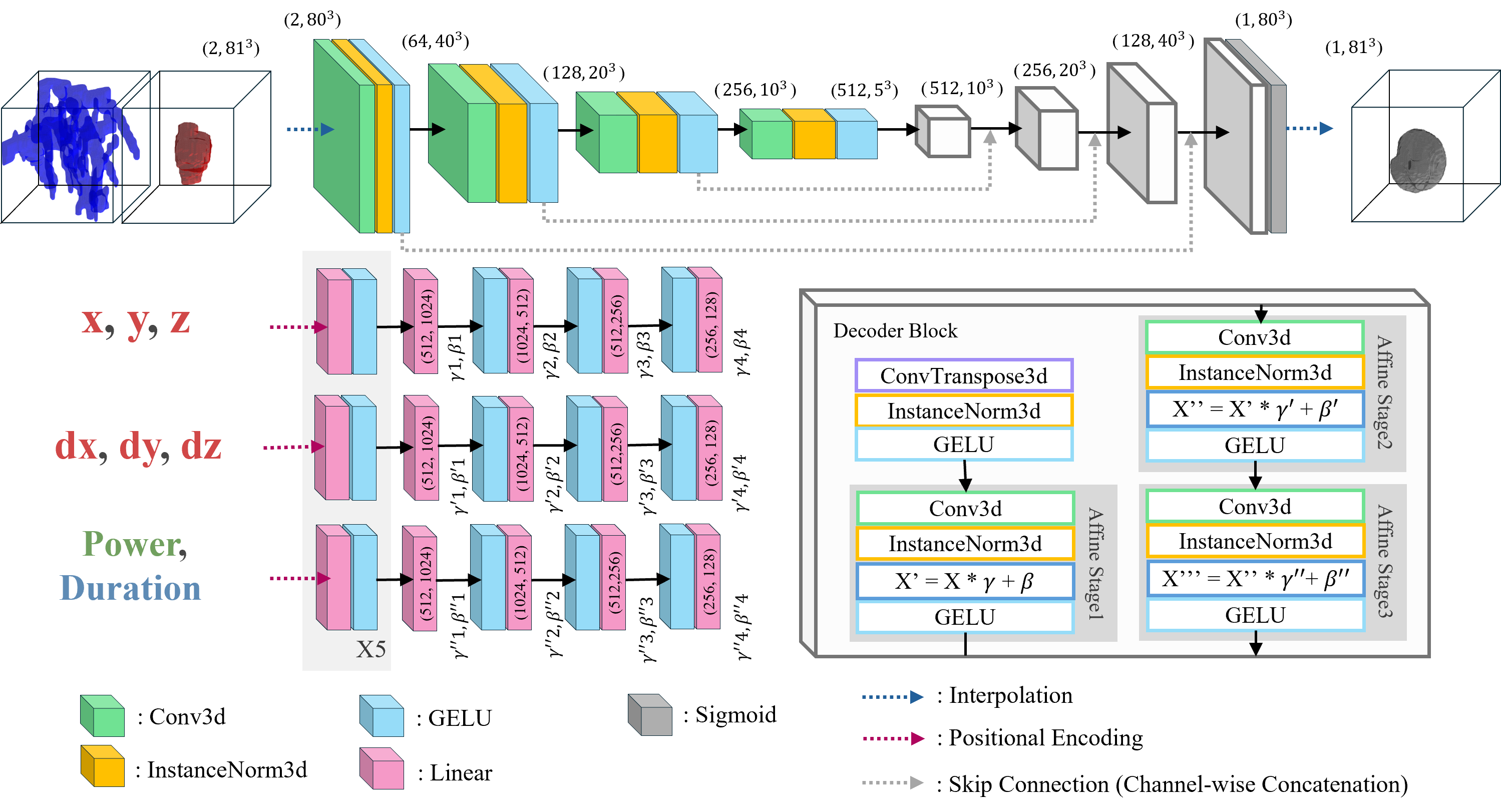}
\caption{Architecture of the neural ablation prediction model. Patient-specific tumor and vessel masks are encoded using a CNN encoder, while antenna configuration and treatment parameters are embedded to generate affine modulation parameters for decoder-stage feature conditioning} \label{fig.1}
\end{figure}

The network architecture shown in Fig.~\ref{fig.1} was designed to approximate the numerical simulation process described above. Accordingly, the model uses the same inputs and output as the numerical simulation and predicts the final ablation region from patient-specific tumor and vessel structures and treatment conditions.

The tumor-vessel volume was processed by a CNN-based encoder. The encoder repeatedly applies 3D CNN layers, instance normalization, and GELU activation to extract high-dimensional features while reducing the spatial resolution. The antenna position, antenna direction, power, and treatment duration were separately passed through positional encoding and linear layers to generate the affine parameters required at each decoder stage, namely the scale parameter $\gamma$ and shift parameter $\beta$.

The extracted features and affine parameters were jointly used in the decoder blocks, allowing the feature maps to be modulated according to the treatment conditions. Each decoder block consists of one upsampling operation and three affine stages. In each affine stage, after a CNN layer, feature modulation was performed by applying $\gamma$ and $\beta$ to the feature map $X$ as $X' = X \cdot \gamma + \beta$ \cite{perez2018film}. The spatial resolution was progressively restored through four decoder blocks, and encoder features were combined with decoder features through skip connections using channel-wise concatenation.

Finally, the output was passed through a sigmoid function to obtain a probability map, which was thresholded at 0.5 to predict the final ablation region.

\subsection{GA-based Treatment Planning}

\subsubsection{3D Body Twin Construction}

We performed GA-based planning to search for patient-specific MWA treatment parameters using the predicted ablation outcome. To account for anatomical constraints along the antenna insertion trajectory, a patient-specific 3D body twin was constructed from the original CT images. The rib region ($\mathrm{HU} \geq 200$) and body region ($\mathrm{HU} \geq -500$) were segmented in 3D Slicer, while major anatomical structures, including the aorta, veins, portal vein, colon, liver, pancreas, and stomach, were segmented using TotalSegmentator \cite{wasserthal2023totalsegmentator}. All segments were resampled to 1-mm voxel spacing and integrated into a single label map to construct the 3D body twin.

\subsubsection{GA-based Optimization Method}
Before optimization, a 5 mm safety margin was added to the target tumor to generate the optimization target mask. The fitness function minimized by the GA was defined as Eq.~\ref{eq:fitness}. The objective function was designed such that the predicted ablation region maximally included the optimization target mask, and Eq.~\ref{eq:tumor_objective} represents the fraction of the target mask that was not ablated. The fraction of the ablated region corresponding to normal tissue, the vertical deflection angle defined as the angle from the horizontal plane, and the antenna insertion trajectory length through the body were treated as soft constraints. Each term was normalized between 0 and 1 and added to the GA fitness function as a weighted penalty term. In contrast, tumor coverage and collision between the antenna insertion trajectory and organs-at-risk were treated as hard constraints that must be satisfied, as defined in Eq.~\ref{eq:constraints}.

\begin{equation}
F(\mathbf{u}) =
Obj(\mathbf{u})
+ 0.3 P_N(\mathbf{u})
+ 0.3 P_{\theta}(\mathbf{u})
+ 0.3 P_L(\mathbf{u})
\label{eq:fitness}
\end{equation}

\begin{equation}
Obj(\mathbf{u}) =
1 -
\frac{\sum_i M_i \hat{A}_i(\mathbf{u})}
{\sum_i M_i + \epsilon}
\label{eq:tumor_objective}
\end{equation}

\begin{equation}
P_N(\mathbf{u}) =
\frac{\sum_i N_i \hat{A}_i(\mathbf{u})}
{\sum_i \hat{A}_i(\mathbf{u}) + \epsilon}
\label{eq:normal_penalty}
\end{equation}

\begin{equation}
\frac{\sum_i \hat{A}_i(\mathbf{u})T_i}
{\sum_i T_i}
= 1,
\qquad
\sum_i R_i(\mathbf{u})O_i = 0.
\label{eq:constraints}
\end{equation}

In Eqs.~\ref{eq:fitness}--\ref{eq:constraints}, 
\(\hat{A}_i(\mathbf{u})\) denotes the thresholded binary ablation mask at voxel \(i\) for the optimization variable \(\mathbf{u}\), obtained by applying a threshold of 0.5 to the model output. 
\(M_i\) and \(T_i\) denote the safety-margin-included target mask and the original tumor mask, respectively.

\(N_i\) denotes the normal tissue mask, defined as the entire domain excluding the target mask, and \(P_N(\mathbf{u})\) penalizes the fraction of the predicted ablation region corresponding to normal tissue. 
\(P_{\theta}(\mathbf{u})\) and \(P_L(\mathbf{u})\) denote the normalized penalties for the vertical deflection angle and antenna insertion trajectory length, respectively. 
The weights of the three penalty terms were all set to 0.3. 
In the hard constraints, complete tumor coverage was enforced by setting the tumor coverage ratio to 1. 
\(O_i\) denotes the organs-at-risk mask excluding the liver, and \(R_i(\mathbf{u})\) denotes the antenna insertion trajectory mask.

The optimization variables \(\mathbf{u}\) included the antenna position, insertion direction, applied power, and treatment duration. During optimization, the patient-specific tumor and vessel masks were fixed, while \(\mathbf{u}\) was updated using the neural ablation prediction model trained in Section 2.2. The GA was run with a population size of 40 for 60 generations, using elitism for the best 4 candidates, tournament selection with a size of 3, blend crossover, and Gaussian mutation with a mutation rate of 0.25 and a mutation scale of 0.1.

\section{Results}

\subsection{Ablation Zone Prediction Performance}
The 60 patient CT datasets used for neural ablation prediction model training and evaluation were divided into training, validation, and test sets with 36, 12, and 12 datasets, respectively. To reduce bias in the data distribution, the datasets were sorted based on tumor size and the number of vessel voxels in the vessel mask, and then evenly distributed across the three sets. The validation set was used for early stopping with a patience of 5 to prevent overfitting during training. Ablation zone prediction performance was evaluated using 1,000 cases randomly sampled from the test set. The model achieved a Dice score of 95.06\% and an IoU of 90.78\%, showing high agreement between the predicted ablation zones and the numerical simulation results.

\subsection{Planning Results}

\subsubsection{Evaluation Metrics}

Planning performance was evaluated using four quantitative metrics: ablation efficiency (AE), complete coverage (CP), organ damage rate (OD), and Insertion path length (IPL). AE measured the proportion of the target mask within the simulated ablation region, while CP measured the coverage of the margin-included target mask by the ablation region. OD represented the proportion of the ablated region overlapping with organs-at-risk, and IPL was defined as the length of the antenna insertion trajectory inside the body mask. Higher AE and CP indicate better target ablation, whereas lower OD and IPL indicate safer and less invasive planning.

\subsubsection{Quantitative Planning Performance}

For planning performance evaluation, 13 patients with liver tumors who were not used for training the prediction model were selected. The selected patients had a maximum tumor diameter of 40 mm or less, with a mean diameter of 21.8 mm (range: 7–36 mm). As a baseline, a clinician specializing in liver tumor MWA procedures was asked to perform conventional clinical planning based on CT images. The same neural ablation prediction model was then applied to the clinician-defined planning conditions to derive the predicted ablation outcomes.

Table~\ref{tab:clinical_validation} presents a quantitative comparison showing that the proposed planning method achieved better mean performance. The mean AE was higher for the proposed planning method than for clinician-defined planning (0.764 vs. 0.495), indicating that the ablation region was more efficiently concentrated within the target mask. CP reached 1.000 for all patients in the proposed planning method, while clinician-defined planning also showed a high mean CP of 0.998, indicating that both methods achieved strong target coverage. In contrast, the mean OD was lower for the proposed planning method than for clinician-defined planning (0.175 vs. 0.389), demonstrating that the proposed method more effectively reduced unnecessary damage to surrounding critical organs. In addition, the mean IPL was 71.2 mm for the proposed planning method and 73.6 mm for clinician-defined planning, indicating that the proposed planning method provided a shorter and less invasive trajectory.

\begin{figure}[H]
\includegraphics[width=\textwidth]{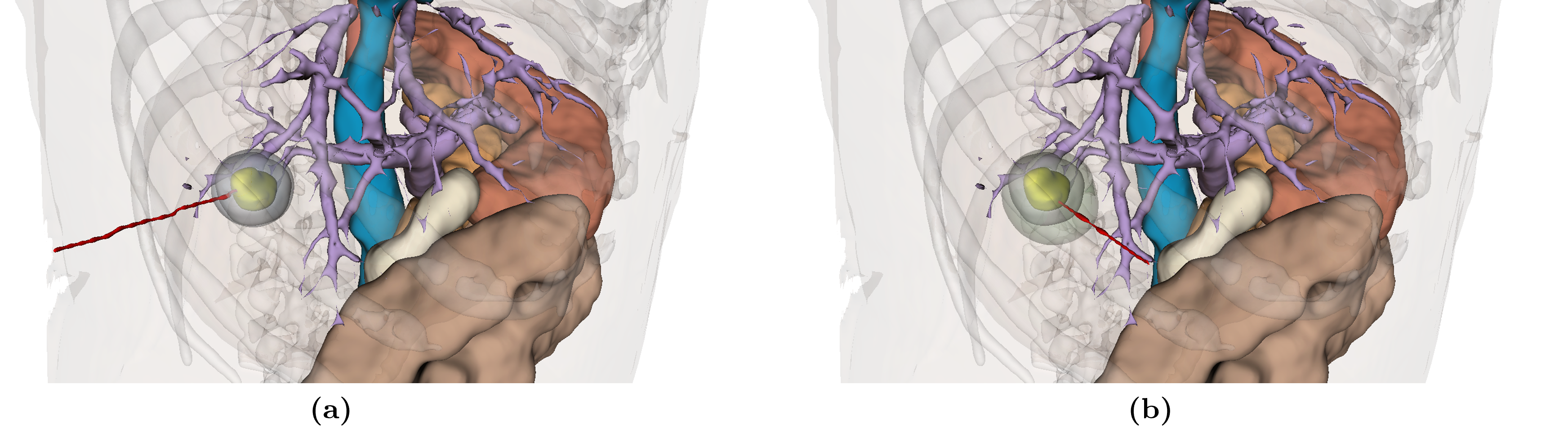}
\caption{Representative treatment planning results comparing the antenna insertion trajectory and predicted ablation region between (a) the proposed method and (b) clinician-defined planning.} \label{fig.2}
\end{figure}

Figure~\ref{fig.2} shows representative planning results for the same patient, comparing the proposed method with clinician-defined planning. The yellow region represents the tumor, and the surrounding boundary indicates the safety margin. The proposed planning result in Fig.~\ref{fig.2}(a) generated an ablation region that closely matched the safety margin, whereas the clinician-defined plan in Fig.~\ref{fig.2}(b) resulted in under-ablation with insufficient coverage of the safety margin.

\begin{table}[t]
\centering
\caption{Quantitative planning performance of the proposed and clinician-defined treatment plans across 13 patients. SD denotes standard deviation.}
\label{tab:clinical_validation}
\begin{tabular}{c c c c c c c c c c}
\toprule
\multirow{2}{*}{Patient} 
& \multirow{2}{*}{\makecell{Tumor diameter \\(mm)}}
& \multicolumn{4}{c}{Proposed planning} 
& \multicolumn{4}{c}{Clinician planning} \\
\cmidrule(lr){3-6} \cmidrule(lr){7-10}
& & AE $\uparrow$ & CP $\uparrow$ & OD $\downarrow$ & IPL $\downarrow$ (mm) & AE $\uparrow$ & CP $\uparrow$ & OD $\downarrow$ & IPL $\downarrow$ (mm) \\
\midrule
1  & 28 & 0.801 & 1.000 & 0.091 & 36.8  & 0.528 & 1.000 & 0.363 & 40.6  \\
2  & 36 & 0.695 & 1.000 & 0.287 & 84.6  & 0.745 & 0.997 & 0.193 & 87.5  \\
3  & 20 & 0.869 & 1.000 & 0.043 & 128.3 & 0.704 & 1.000 & 0.147 & 133.9 \\
4  & 27 & 0.873 & 1.000 & 0.096 & 65.6  & 0.537 & 1.000 & 0.333 & 70.5  \\
5  & 28 & 0.539 & 1.000 & 0.215 & 58.4  & 0.394 & 0.996 & 0.340 & 73.4  \\
6  & 20 & 0.636 & 1.000 & 0.276 & 67.8  & 0.535 & 1.000 & 0.379 & 64.0  \\
7  & 14 & 0.836 & 1.000 & 0.164 & 81.7  & 0.361 & 1.000 & 0.639 & 79.7  \\
8  & 7  & 0.843 & 1.000 & 0.157 & 43.0  & 0.200 & 1.000 & 0.676 & 44.1  \\
9  & 11 & 0.820 & 1.000 & 0.106 & 60.0  & 0.364 & 1.000 & 0.406 & 39.6  \\
10 & 20 & 0.817 & 1.000 & 0.183 & 49.1  & 0.487 & 0.999 & 0.507 & 62.4  \\
11 & 23 & 0.786 & 1.000 & 0.120 & 76.1  & 0.672 & 1.000 & 0.240 & 85.0  \\
12 & 27 & 0.789 & 1.000 & 0.155 & 60.0  & 0.318 & 1.000 & 0.415 & 52.4  \\
13 & 23 & 0.620 & 1.000 & 0.379 & 114.1 & 0.587 & 0.986 & 0.413 & 124.0 \\
\midrule
Mean & 21.8 & \textbf{0.764} & \textbf{1.000} & \textbf{0.175} & \textbf{71.2} & \textbf{0.495} & \textbf{0.998} & \textbf{0.389} & \textbf{73.6} \\
SD & 7.9 & 0.106 & 0.000 & 0.094 & 26.4 & 0.162 & 0.004 & 0.154 & 29.3 \\
\bottomrule
\end{tabular}
\end{table}

\subsubsection{Clinical Feasibility Assessment}

The clinical applicability of the proposed planning method was evaluated by two MWA specialists using four categories: optimal, acceptable, indeterminate, and not feasible. Plans rated as optimal or acceptable were considered clinically applicable. As shown in Table~\ref{tab:clinical_validation_detail}, all antenna insertion trajectories were rated as optimal or acceptable by both clinicians, indicating clinically acceptable trajectory generation. In contrast, treatment duration and power settings showed greater inter-clinician variability, with one case rated as indeterminate due to possible insufficient ablation, suggesting the need for more quantitative criteria for treatment parameter selection.

\begin{table}[t]
\centering
\caption{Clinician assessment of the feasibility of antenna insertion trajectory and time/power settings.}
\label{tab:clinical_validation_detail}
\begin{tabular}{c c c c c c}
\toprule
Evaluation item & Evaluator & Optimal & Acceptable & Indeterminate & Not feasible \\
\midrule
\multirow{2}{*}{Antenna insertion trajectory} 
& Clinician 1 & 9 & 4 & -- & -- \\
& Clinician 2 & 10 & 3 & -- & -- \\
\midrule
\multirow{2}{*}{Time/power setting} 
& Clinician 1 & 13 & -- & -- & -- \\
& Clinician 2 & 6 & 6 & 1 & -- \\
\bottomrule
\end{tabular}
\end{table}

\subsection{Planning Time Analysis}

Using the neural ablation prediction model as the forward model substantially reduced planning time. The model inference time was 16.3 ms per case, compared with an average of 10.8 s for numerical simulation. The neural-network-based planning required an average of 42.7 s, including 1,664 forward evaluations taking 32.9 s. Performing the same number of evaluations with numerical simulation would require approximately 17,971 s (5.0 h), indicating that the proposed framework enabled about 420-fold faster planning. All measurements were conducted on a GPU-enabled desktop with an AMD Ryzen 9 5900X CPU and an NVIDIA RTX 3090 GPU.

\section{Conclusion}

In this study, we proposed a patient-specific MWA planning framework that combines a neural network-based ablation prediction model with a genetic algorithm. Trained on multiphysics simulation data, the model rapidly predicts ablation regions by incorporating tumor and vessel structures, antenna configuration, and treatment conditions. Compared with clinician-defined planning, the proposed method improved ablation efficiency while reducing organ damage and antenna insertion trajectory length, with most plans judged clinically applicable by MWA specialists. By enabling approximately 420-fold faster planning than numerical-simulation-based planning, the proposed framework shows its potential as a fast digital twin-based system for quantitative, automated, and personalized MWA treatment planning.


\bibliographystyle{splncs04}
\bibliography{paper}

\end{document}